\documentclass[%
 reprint,
superscriptaddress,
footinbib,
 amsmath,amssymb,
 prl
]{revtex4-2}
\usepackage{amsmath}
 \usepackage{color} 
\usepackage{graphicx}% Include figure files
\usepackage{dcolumn}% Align table columns on decimal point
\usepackage{bm}% bold math
\usepackage{subfigure}
\usepackage{multirow}
\usepackage{hyperref}% add hypertext capabilities
\usepackage[mathlines]{lineno}% Enable numbering of text and display math
\begin{document}

\preprint{APS/123-QED}

\title{Observation of the electromagnetic radiative decays of the  \boldmath{$\Lambda(1520)$} and 
\boldmath{$\Lambda(1670)$} to \boldmath{$\gamma\Sigma^0$} }
%and search for \boldmath{$\Lambda(1670)\to\gamma\Lambda$}}

\author{
\begin{small}
  \begin{center}
%% Saved at => 2025-03-17
M.~Ablikim$^{1}$, M.~N.~Achasov$^{4,c}$, P.~Adlarson$^{77}$, X.~C.~Ai$^{82}$, R.~Aliberti$^{36}$, A.~Amoroso$^{76A,76C}$, Q.~An$^{73,59,a}$, Y.~Bai$^{58}$, O.~Bakina$^{37}$, Y.~Ban$^{47,h}$
% ...
\\
\vspace{0.2cm}
(BESIII Collaboration)\\
\vspace{0.2cm} {\it
$^{1}$ Institute of High Energy Physics, Beijing 100049, People's Republic of China\\
% ...
$^{4}$ Budker Institute of Nuclear Physics SB RAS (BINP), Novosibirsk 630090, Russia\\
% ...
$^{36}$ Johannes Gutenberg University of Mainz, Johann-Joachim-Becher-Weg 45, D-55099 Mainz, Germany\\
$^{37}$ Joint Institute for Nuclear Research, 141980 Dubna, Moscow region, Russia\\
% ...
$^{47}$ Peking University, Beijing 100871, People's Republic of China\\
% ...
$^{58}$ Southeast University, Nanjing 211100, People's Republic of China\\
$^{59}$ State Key Laboratory of Particle Detection and Electronics, Beijing 100049, Hefei 230026, People's Republic of China\\
% ...
$^{73}$ University of Science and Technology of China, Hefei 230026, People's Republic of China\\
% ...
$^{76}$ University of Turin and INFN, (A)University of Turin, I-10125, Turin, Italy; (B)University of Eastern Piedmont, I-15121, Alessandria, Italy; (C)INFN, I-10125, Turin, Italy\\
$^{77}$ Uppsala University, Box 516, SE-75120 Uppsala, Sweden\\
% ...
$^{82}$ Zhengzhou University, Zhengzhou 450001, People's Republic of China\\

\vspace{0.2cm}
$^{a}$ Deceased\\
% ...
$^{c}$ Also at the Novosibirsk State University, Novosibirsk, 630090, Russia\\
% ...
$^{h}$ Also at State Key Laboratory of Nuclear Physics and Technology, Peking University, Beijing 100871, People's Republic of China\\
% ...

}
%% ends here %%

\end{center}
\end{small}
}

\date{\today}
\begin{abstract}
Using $(10087\pm 44)\times10^6$ $J/\psi$ events collected with the BESIII detector, we report the first observation of the electromagnetic radiative decays of the $\Lambda(1520)$ and $\Lambda(1670)$ to $\gamma\Sigma^0$, with a statistical significance of $16.6\sigma$ and $23.5\sigma$, respectively. The ratio of the branching fractions $\frac{\mathcal{B}(\Lambda(1520)\to\gamma\Lambda)}{\mathcal{B}(\Lambda(1520)\to\gamma\Sigma^0)}$ is determined to be $2.88\pm0.27(\text{stat.})\pm0.21(\text{syst.})$, which is in good agreement with flavor SU(3) symmetry. The branching fraction of $\Lambda(1520)\to\gamma\Sigma^0$ is measured to be $\mathcal{B}(\Lambda(1520)\to\gamma\Sigma^0)=(2.95\pm0.28(\text{stat.})\pm0.56(\text{syst.}))\times 10^{-3}$, corresponding to a partial width of $\Gamma(\Lambda(1520)\to\gamma\Sigma^0)=(47.2\pm4.5(\text{stat.})\pm9.0(\text{syst.}))$ keV, which is inconsistent with predictions from the relativized constituent quark model and the Algebraic model. Additionally, we observe a clear resonant structure in the $\gamma\Sigma^0$ mass spectrum around 1.67 GeV/$c^2$, attributed to the $\Lambda(1670)$. The product branching fraction $\mathcal{B}(J/\psi\to\bar\Lambda\Lambda(1670)+c.c.)\times\mathcal{B}(\Lambda(1670)\to\gamma\Sigma^0)$ is measured for the first time as $(5.39\pm0.29(\text{stat.})\pm 0.44(\text{syst.}))\times 10^{-6}$. However, no corresponding structure is seen in the $\gamma\Lambda$ mass spectrum, so an upper limit on the product branching fraction $\mathcal{B}(J/\psi\to\bar\Lambda\Lambda(1670)+c.c.)\times\mathcal{B}(\Lambda(1670)\to\gamma\Lambda)$ is determined to be $5.97\times10^{-7}$ at the 90\% confidence level.
\end{abstract}
\maketitle

%%%%%%%%%%%%%%%%%%%

Hyperons decay radiatively to lower-energy states by emitting a photon. The emitted photon carries information about the properties of the initial and final hyperon states, making it a unique probe for exploring the underlying quark structure and interactions within hyperons.  This has attracted attention from both experimental and theoretical perspectives.
While the importance of these decays has been stressed for over half a century, 
it has remained a significant challenge to find a consistent picture between the theoretical descriptions and the experimental results for hyperon radiative decays.

Some different theoretical approaches~\cite{Bertini,Isgur1,Isgur2,Yu:2006sc}, both at the baryon and at the quark level, have
been employed to investigate hyperon radiative decays, and significant theoretical progress has been made. However, a completely satisfactory description is still lacking. The experimental situation has improved significantly over the past few years with the advent of the BESIII experiment. Taking advantage of the high production rate of hyperons in $J/\psi$ decays,  several radiative hyperon decays, in particular for the octet ground states, have been measured by BESIII for the first time~\cite{SigmaP,LambdaN}.

However, experimental results for radiative decays of excited hyperons are still very sparse. Even for the narrow $\Lambda(1520)$, the most recent data on $\Lambda(1520)\rightarrow \gamma \Lambda$ dates back twenty years~\cite{CLAS}, and the decay $\Lambda(1520)\rightarrow \gamma \Sigma^0$ has not yet been observed. 
These decays are of particular interest as they offer further insight into the internal structure of
the  $\Lambda(1520)$, which has inspired  considerable  theoretical speculation.  Assuming it is a flavor SU(3) singlet,  the predicted ratio of  decay widths between $\Lambda (1520)\rightarrow \gamma \Lambda$ and $\Lambda (1520)\rightarrow \gamma \Sigma^0$ is approximately 2.5~\cite{Landsberg}.  Theoretical predictions  from  models such as the $\chi$QM, NRQM, RCQM, MIT Bag, Chiral Bag and Algebraic models, as summarized in Table~\ref{pre_exp},  show significant variations in the expected rates for these two processes.  Therefore, more precise measurements of the branching fractions of these excited hyperon decays are not only crucial for validating theoretical models but also provide valuable insights into the interactions among confined light quarks.

In this Letter, we report the first observation of the electromagnetic radiative transition $\Lambda(1520)\to\gamma\Sigma^0$ with a sample of $(10087\pm 44)\times 10^6$ $J/\psi$ events~\cite{jpsidata}. The measurement of the ratio 
$\frac{\mathcal{B}(\Lambda(1520)\to\gamma\Lambda)}{\mathcal{B}(\Lambda(1520)\to\gamma\Sigma^0)}$, 
along with the corresponding partial width of $\Lambda(1520)$, is also presented.  In addition, a clear resonant peak around 1.67 GeV/$c^2$, which is attributed to the $\Lambda(1670)$,  in the $\gamma\Sigma^0$ mass spectrum is observed for the first time, but no evident signal of $\Lambda(1670)\rightarrow\gamma\Lambda$ is seen.  The inclusion of charge-conjugate processes is implied.

\begin{table}[htp]
    \caption{Theoretical predictions and experimental measurements of the partial widths of $\Lambda (1520)\rightarrow \gamma \Lambda$ and $\Lambda (1520)\rightarrow \gamma \Sigma^0$ (in keV).}
    \centering{
    \begin{tabular}{l|c|c}
    \hline
    \hline  Decay &  $\Lambda({1520})\rightarrow \gamma \Lambda$ &  $\Lambda(1520)\rightarrow \gamma \Sigma^0$ \\ 
    \hline
    {$\chi$QM~\cite{Yu:2006sc}}& 134 & 92\\
    NRQM & 156~\cite{QM3},96~\cite{QM4} & 55~\cite{QM3}, 74~\cite{QM4} \\
    RCQM~\cite{RCQM}& 215 & 293 \\
    MIT\;Bag~\cite{QM3}& 46 & 17\\
    Chiral\;Bag~\cite{CBM5}& 32 & 49 \\
    Algebraic\; model~\cite{AM} &85.1 & 180.4 \\
    \hline
    CLAS Collaboration~\cite{CLAS} & $167\pm43^{+26}_{-12}$ &- \\
\hline    \hline
    \end{tabular}}
    \label{pre_exp}
\end{table}

%%%%%%%%%%%%%%%%%%%%

The design and performance of the BESIII detector are described in detail in Refs.~\cite{BESIIIdetector,BESIIwhite}. Simulated samples, including both inclusive and exclusive events produced with the {\sc Geant}4-based~\cite{Geant4} Monte Carlo (MC) package, which encompasses the geometric description of the BESIII detector and its response, are utilized to determine the detection efficiency and estimate backgrounds. The processes $J/\psi\to\bar\Lambda\Lambda(1520/1670)$ and the subsequent electromagnetic transitions $\Lambda(1520/1670)\rightarrow\gamma\Lambda(\Sigma^0)$ are simulated using the phase space (PHSP) model implemented in the {\sc EvtGen} MC event generator~\cite{EVGEN1,EVGEN2}.
%The production of the $\Lambda(1520/1670)$ resonances from the decay $J/\psi\to\bar\Lambda\Lambda(1520/1670)$, along with the electromagnetic transitions $\Lambda(1520/1670)\rightarrow\gamma\Lambda(\Sigma^0)$, are simulated using the phase space (PHSP) model from the MC event generator {\sc EvtGen}~\cite{EVGEN1,EVGEN2}.

In the analysis of  both $\Lambda(1520/1670)\to\gamma\Lambda$ (Mode I) and $\Lambda(1520/1670)\to\gamma\Sigma^0$ (Mode II),  the $\Sigma^0$ and $\Lambda$ are reconstructed with their decay modes of $\Sigma^0\rightarrow\gamma\Lambda~(\Lambda\to p \pi^-)$  and $\Lambda\to p\pi^-$, respectively.
The candidate charged tracks and electromagnetic showers are required to satisfy the following common selection criteria. (i) Charged tracks are reconstructed using the tracking information from the multi-layer drift chamber (MDC). The distance of the closest approach of each charged track to the $e^+e^-$ interaction point is required to be within $\pm20$ cm along the beam direction and within 10~cm in the plane perpendicular to the beam direction. The polar angle $\theta$ between the direction of a charged track and that of the beam must satisfy $|\text{cos}\theta|<0.93$ for an effective measurement in the active volume of the MDC. (ii) Electromagnetic showers are reconstructed from clusters of deposited energy in the electromagnetic calorimeter (EMC). The shower energies of photon candidates must exceed \mbox{25 MeV} in the barrel region ($|\text{cos}\theta|<0.80$) or 50 MeV in the endcap regions ($0.86 <|\text{cos}\theta|< 0.92$). Showers located in the transition regions between the barrel and the endcap regions are excluded. To avoid showers caused by charged particles, a photon candidate has to be separated by at least $10^\circ$ from any charged track. To suppress electronic noise and unrelated energy deposits, the EMC time $t$ of the photon candidates must fall within the range $0\le t\le 700$ ns. (iii) To reconstruct the decays $\Lambda\to p \pi^-$ and $\bar\Lambda\to\bar{p}\pi^+$, we loop over all the combinations of positive and negative charged track pairs and require that at least one $( p \pi^-)(\bar{p}\pi^+)$ track hypothesis successfully passes the vertex finding algorithm~\cite{vertexfit} of $\Lambda$ and 
$\bar{\Lambda}$. If multiple $p\pi^-$ combinations satisfy the vertex fit requirement, the one with the mass closest to 
$M_{\Lambda}$  is chosen, where  $M_{\Lambda}$  is the nominal mass of $\Lambda$~\cite{PDG}.
The $p \pi^-$ and $\bar{p}\pi^+$ pairs are accepted as $\Lambda$ and $\bar\Lambda$ candidates if their invariant masses satisfy $|M_{p\pi^-}-M_{\Lambda}|<5$ MeV/$c^2$ and $|M_{\bar{p}\pi^+}-M_{\Lambda}|<5$ MeV/$c^2$, respectively.

\begin{figure}[!htbp]
	    \centering
        \includegraphics[width=0.45\textwidth]{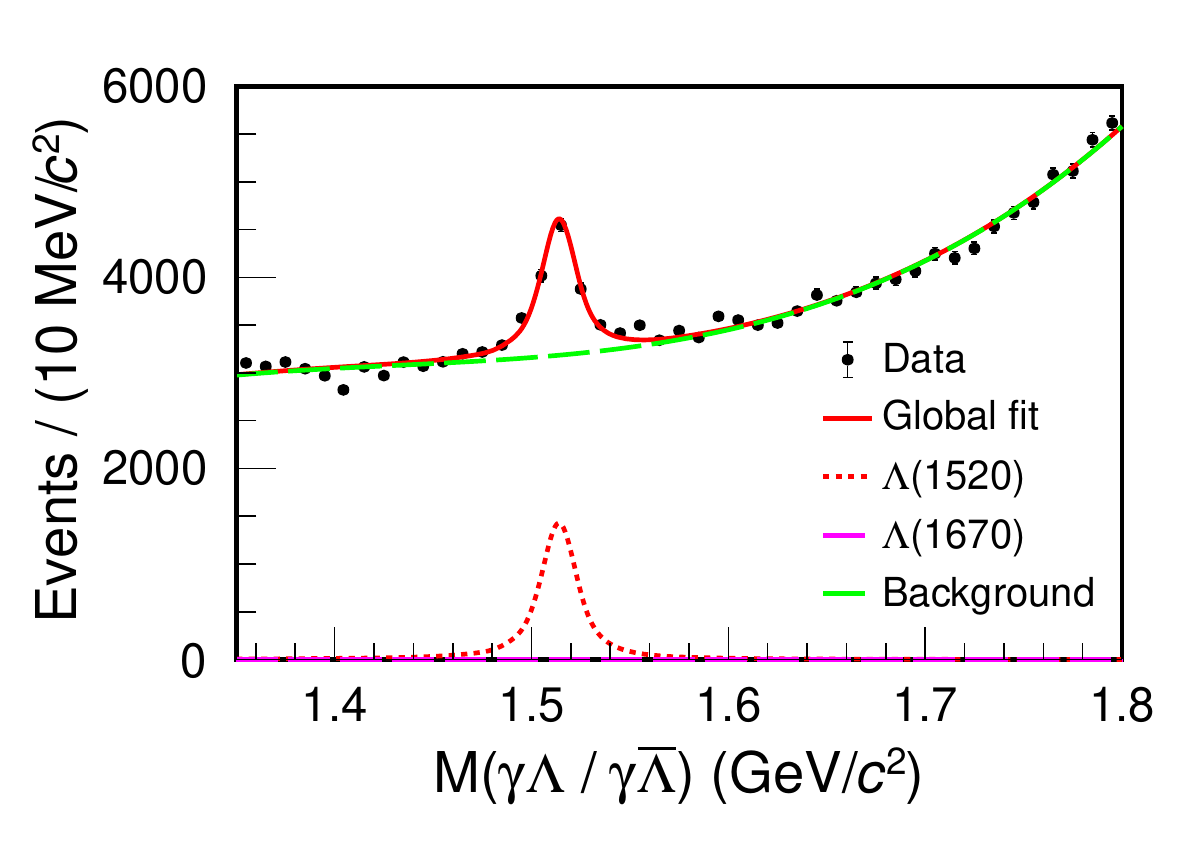}
	    \caption{Fit result of $M(\gamma\Lambda/\gamma\bar\Lambda)$. The black dots with error bars are data, the red curve is the global fit, the red dotted-curve is the $\Lambda(1520)$ signal, the pink long-dashed curve is the $\Lambda(1670)$ signal and the green long-dashed curve is the fitted background shape. }\label{combinefit_gamlam}
    \end{figure}

For Mode I, to further suppress backgrounds and improve mass resolution, a four-constraint energy momentum conservation (4C) kinematic fit is performed with the hypothesis of $\gamma \Lambda\bar{\Lambda}$ and events with $\chi^2_{\rm 4C}<30$ are retained. In addition, $|M_{\gamma\bar{\Lambda}}-M_{\Sigma^0}|>0.02$ GeV/$c^2$ and $|M_{\gamma{\Lambda}}-M_{\Sigma^0}|>0.02$ GeV/$c^2$ are required in the further analysis for $J/\psi\to\bar\Lambda\Lambda(1520)$ and $J/\psi\to\Lambda\bar{\Lambda}(1520)$, respectively, which effectively suppresses the $J/\psi\to\bar\Lambda\Sigma^0$ background. Following these requirements, the invariant mass spectrum of $\gamma\Lambda~(\gamma\bar{\Lambda})$ is displayed in Fig.~\ref{combinefit_gamlam}, where the prominent peak of $\Lambda(1520)$ is clearly observed while there is no significant $\Lambda(1670)$ signal. Potential background events are investigated using an inclusive MC sample of 10 billion $J/\psi$ events generated with the {\sc EvtGen}~\cite{EVGEN1,EVGEN2} and {\sc LundCharm}~\cite{lundcharm1,lundcharm2}. Employing the same selection criteria, it is indicated that the main background originates from $J/\psi\to\gamma\Lambda\bar\Lambda$, with no evident peak in the $\Lambda(1520/1670)$ mass region.

\begin{figure}[!htbp]
	    \centering
        \includegraphics[width=0.45\textwidth]{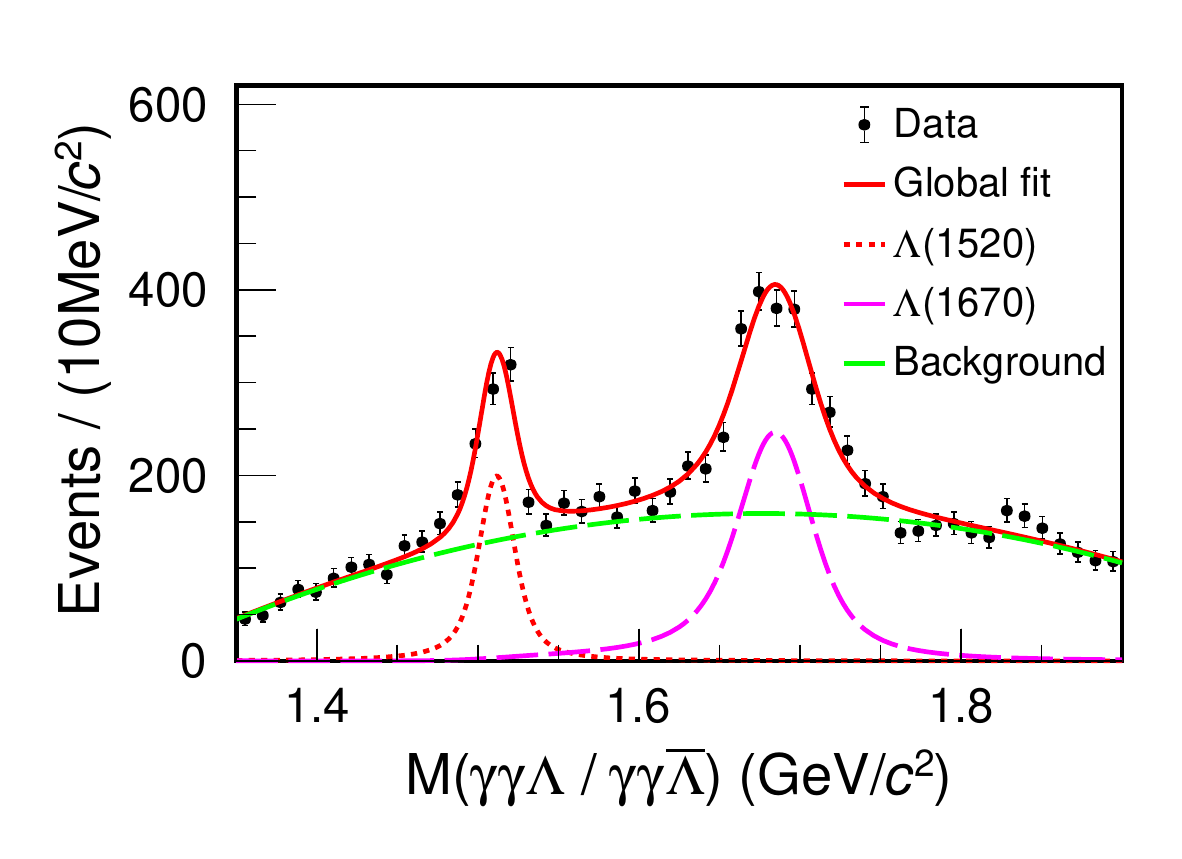}
	    \caption{Fit result of $M(\gamma\gamma\Lambda/\gamma\gamma\bar\Lambda)$. The black dots with error bars are data, the red curve is the global fit, the red dotted-curve is the $\Lambda(1520)$ signal, the pink long-dashed curve is the $\Lambda(1670)$ signal and the green long-dashed curve is from the fitted background shape.}\label{combinefit_gamgamlam}
    \end{figure}

For Mode II,
% to further suppress backgrounds and improve mass resolution, 
a 4C kinematic fit is performed under the hypothesis of $\gamma\gamma\Lambda\bar{\Lambda}$, retaining those events with $\chi^2_{\rm 4C}<15$. 
To effectively suppress background events from $J/\psi\to\bar\Lambda\Sigma^0\pi^0$, we implement a sequential analysis procedure involving kinematic fits and selection criteria. First, a one-constraint (1C) kinematic fit is applied to the $ \gamma\Lambda $ hypothesis with the invariant mass constrained to the $ \Sigma^0 $ nominal mass, enabling extraction of the $ \bar{\Lambda}\Sigma^0 $ squared recoil mass $(M_{\bar{\Lambda}\Sigma^0}^{\text{rec}})^2$ and significantly reducing this background. Subsequently, a two-constraint (2C) kinematic fit is performed on the complete $ \gamma, \pi^0 (\to \gamma\gamma_{\text{miss}}), \Lambda, \bar{\Lambda} $ hypothesis, incorporating energy-momentum conservation and requiring the $\gamma\gamma_{\text{miss}}$ system to match the $\pi^0$ mass. The combined application of selection criteria $ (M_{\bar{\Lambda}\Sigma^0}^{\text{rec}})^2 < 0.01 \, (\text{GeV}/c^2)^2 $ and $ \chi^{2}_{\text{2C}} > 5 $ provides additional background suppression, substantially enhancing the purity of the signal events through this two-stage approach.
To suppress background events with  $\pi^0$ in the final state, we require $|M_{\gamma\gamma}-M_{\pi^0}|>0.03$ GeV/$c^2$.  The events with both $\gamma\Lambda$ and $\gamma\bar{\Lambda}$ in the $\Sigma^0$ mass regions, 
% $|M_{\gamma_{\rm low}\Lambda}-M_{\Sigma^0}|<0.01$ GeV/$c^2$, 
are rejected for the background events from  $J/\psi\to\Sigma^0\bar\Sigma^0$.
Since the photon from $\Lambda(1520/1670)$ generally has higher energy than that from $\Sigma^0$ decay, the lower-energy photon is assumed to originate from $\Sigma^0$, and a requirement of $|M_{\gamma_{\rm low}\Lambda} - M_{\Sigma^0}| < 0.01$ GeV/$c^2$ is applied to select the $\Sigma^0$ signal. After these requirements, the invariant mass spectrum of $\gamma\gamma\Lambda~(\gamma\gamma\bar{\Lambda})$ is presented in Fig.~\ref{combinefit_gamgamlam}, where the prominent peaks of $\Lambda(1520)$ and $\Lambda(1670)$ are clearly observed. Possible background sources are investigated with an inclusive MC sample of 10 billion $J/\psi$ events. Employing the same selection criteria, it is indicated that the main background originates from $J/\psi\to\gamma\bar\Lambda\Sigma^0$, with no peaking structure observed in the $\Lambda(1520/1670)$ signal region.

The signal yields of the exclusive channels are derived from unbinned maximum likelihood fits applied to the mass spectra of selected candidates for $\Lambda(1520/1670) \to \gamma \Lambda$ and $\Lambda(1520/1670) \to \gamma \Sigma^0$. In these fits, the signal components are modeled using MC simulated shapes convolved with a Gaussian function to account for discrepancies in mass resolution between data and MC simulation. The background is represented by the third-order and second-order Chebyshev polynomial functions for Mode I and Mode II, respectively, with free background yield. The fits, as illustrated in Figs.~\ref{combinefit_gamlam} and \ref{combinefit_gamgamlam}, yield $3738 \pm 194$ events for $\Lambda(1520) \to \gamma \Lambda$, $636 \pm 49$ events for $\Lambda(1520) \to \gamma \Sigma^0$ with a statistical significance of $16.6\sigma$, and $1627 \pm 86$ events for $\Lambda(1670) \to \gamma \Sigma^0$ with a statistical significance of $23.5\sigma$.  With the detection efficiencies obtained from the MC simulations, the ratio of the branching fractions is calculated with
\begin{equation}
\mathcal{R} = \frac{\mathcal{B}(\Lambda(1520/1670)\to\gamma\Lambda)}{\mathcal{B}(\Lambda(1520/1670)\to\gamma\Sigma^0)} = \frac{N_{\gamma\Lambda}^{\rm obs}}{N_{\gamma\Sigma^0}^{\rm obs}} \cdot \frac{\epsilon_{\gamma\Sigma^0}}{\epsilon_{\gamma\Lambda}},
\end{equation}
where $N_{\gamma\Lambda}^{\rm obs}$ and $N_{\gamma\Sigma^0}^{\rm obs}$ are the signal yields (or the upper limit at the 90\% confidence level), and $\epsilon_{\gamma\Sigma^0}$ and $\epsilon_{\gamma\Lambda}$, are the corresponding detection efficiencies. With the
numbers given in Table~\ref{SumTable1_Lambda1520_combineFit}, $\mathcal{R}$ for $\Lambda(1520)$ decay is determined to be $2.88\pm0.27$, where the uncertainty is statistical only.
The product branching fraction is calculated with
\begin{equation}
\begin{aligned}
    & \mathcal{B}(J/\psi\to\bar\Lambda\Lambda(1670))\times\mathcal{B}(\Lambda(1670)\to\gamma\Lambda(\Sigma^0)) \\
    & = \frac{N_{\gamma\Sigma^0}^{\rm obs}}{N_{J/\psi} \times \mathcal{B}^2(\Lambda\rightarrow p \pi) \times \epsilon_{\gamma\Sigma^0}},
\end{aligned}
\end{equation}
where $N_{J/\psi}$ is the total number of $J/\psi$ events and $\mathcal{B}(\Lambda\rightarrow p \pi)$ is cited from the Particle Data Group (PDG)~\cite{PDG}. With the
numbers given in Table~\ref{SumTable1_Lambda1520_combineFit}, $\mathcal{B}(J/\psi\to\bar\Lambda\Lambda(1670))\times\mathcal{B}(\Lambda(1670)\to\gamma\Sigma^0)$ is determined to be $(5.39\pm0.29)\times 10^{-6}$, where the uncertainty is statistical only.

\begin{table}%[t]
\renewcommand{\arraystretch}{1.5}
\caption{ Signal yields, detection efficiencies and the $\mathcal{R}$ of each signal decay. The first
uncertainties are statistical and the second systematic.
}\label{SumTable1_Lambda1520_combineFit}
\centering
\begin{tabular}{l|c|c|c}
   \hline
   Decay   &$N^{\rm obs}$&$\epsilon_{\gamma\Lambda/\gamma\Sigma^0}$& $\mathcal{R}$ \\
   \hline\hline
   {$\Lambda(1520)\to\gamma\Lambda$}&$3738\pm194$&18.91\%&\multirow{2}{*}{$2.88\pm0.27\pm0.21$} \\
    \cline{1-3}
   {$\Lambda(1520)\to\gamma\Sigma^0$}&$636\pm49$&9.28\%&~\\
  \hline 
      {$\Lambda(1670)\to\gamma\Lambda$}& $<412$&16.66\%&\multirow{2}{*}{$<0.11$} \\
    \cline{1-3}
   {$\Lambda(1670)\to\gamma\Sigma^0$}&$1627\pm86$&7.28\%&~\\
   
    \hline
\end{tabular}
\end{table}

Since no significant signal is evident for the decay $\Lambda(1670)\to\gamma\Lambda$, as illustrated in Fig.~\ref{combinefit_gamlam},  we set the upper limit on the signal yield at the 90\% confidence level with the Bayesian approach~\cite{bayesian}. 
In order to determine the likelihood distribution, a series of unbinned extended maximum likelihood fits are performed to the mass spectrum of $\gamma\Lambda$ scanning along the magnitude of the $\Lambda(1670)$ signal. 
In the fit, the line shape of the $\Lambda(1670)$ signal is determined by MC simulation, and the background is represented with a third-order Chebychev polynomial function.  The resultant likelihood distribution is taken as the probability density function directly. The upper limit on the number of signal events at the 90\% confidence level is defined as $N_{\rm UL}$, corresponding to the number of events at 90\% of the integral of the probability density function of likelihood distribution.  To account for the additive systematic uncertainties from the mass spectrum fit, the alternative fits are performed with different fit ranges and background shapes and the maximum one,  $N_{\rm UL}$=407, is used to evaluate the upper limit on the branching fraction of the interest decay.

The final upper limit on the product branching fraction also considers the multiplicative systematic uncertainty, which is accounted for by convolving the likelihood distribution $\mathcal{L}(N)$ to obtain the smeared likelihood $\mathcal{L}'(N)$,
\begin{equation}\label{eqBr0_BF2_3}
  \mathcal{L}'(N)={\displaystyle{{\int}^{1}_{0}} \mathcal{L}  ( \frac{S}{\hat{S}}  N)  {\rm exp}  \left[ - \frac{\left(S - \hat{S}\right)}{2 {\sigma}^2_{S}}  \right]  d S  }.
\end{equation}
In this expression $\hat{S}$ is the nominal efficiency,
$\sigma_{S}$ is its absolute systematical uncertainty. Following this procedure, the upper limit on the number of $\Lambda(1670)\to\gamma\Lambda$ signal events at the $90\%$ confidence level is set to be $412$, and the corresponding upper limit of the product branching fraction is $\mathcal{B}(J/\psi\to\bar\Lambda\Lambda(1670),\Lambda(1670)\to\gamma\Lambda)<5.97 \times {10^{ - 7}}$, which yields an upper limit of $\mathcal{R}$ for $\Lambda(1670)$ decay of $0.11$.

Sources of systematic uncertainties and their corresponding
contributions to the measurements of the branching fractions
are summarized in Table~\ref{totSys_combine}.

The uncertainty from the total number of $J/\psi$ events, as reported in Ref.~\cite{jpsidata}, is 0.4\%. The photon detection efficiency is evaluated using the control sample of $J/\psi\to\rho\pi$~\cite{BRgammaefficicy}, resulting in an uncertainty of 1.0\% per photon. Using the clean control sample $J/\psi\to\bar{p}K^{+}\Lambda$, the momentum-dependent $\Lambda$ reconstruction efficiency combining the tracking efficiency has been studied in a given cos$\theta$ range. The difference on efficiency between data and MC simulation is taken as the corresponding systematic uncertainty.
The uncertainty associated with the kinematic fit comes
from the inconsistency between data and MC simulation of
the track parameters and the error matrices. The  helix parameters of the charged tracks, and their errors of the MC simulation are then corrected to 
improve the consistency between data and MC simulation~\cite{helixcorrection}; the difference of the signal efficiencies before and after this correction is taken as the systematic uncertainty.

The uncertainty associated with the signal model is assessed by replacing the PHSP model with helicity amplitudes derived from Ref.~\cite{J2BB3} for $J/\psi\to\bar\Lambda\Lambda(1520)$ and Ref.~\cite{J2BB1} for $J/\psi\to\bar\Lambda\Lambda(1670)$. The difference in detection efficiency is attributed to systematic uncertainty. Additionally, the fitting related uncertainty, including fit range and background shape, are determined
by varying the background functions and the fit ranges.
The uncertainties arising from the veto of   $\Lambda\bar{\Sigma}^0\pi^0$ events are estimated by varying the requirements of $(M^{\rm rec}_{\Lambda \Sigma})^{2}$ and $\chi^{2}_{\rm 2C}$.  And the maximum changes of the results for   $\Lambda(1520)\to\gamma\Sigma^0$
and $\Lambda(1670)\to\gamma\Sigma^0$, respectively, are assigned as the systematic uncertainties.

\begin{table}
\renewcommand{\arraystretch}{1.5}
\centering
 \caption{Relative systematic uncertainties (in \%) for the branching fraction measurements. Mode I(a) and Mode II(a) correspond to $\Lambda(1520)\to\gamma\Lambda$ and $\Lambda(1520)\to\gamma\Sigma^0$. Mode I(b) and Mode II(b) correspond to $\Lambda(1670)\to\gamma\Lambda$ and $\Lambda(1670)\to\gamma\Sigma^0$, respectively. The items with $*$ are common uncertainties of Mode I(a) and Mode II(a), and the other items are independent uncertainties. In Mode I(b), the additive systematic uncertainties associated with the fitting procedures donated by ``$\dots$" are excluded, as they have already been accounted for in the upper limit described above.}\label{totSys_combine}
 \vspace{0.2cm}
\noindent
\footnotesize
 \begin{tabular}{c|c|c|c|c}
 %\begin{tabular}[0.618\textwidth]{ccccc}
 \hline
 \multirow{2}{*}{Source} & \multicolumn{2}{c|}{$\mathcal{R}$}  & \multirow{2}{*}{Mode I(b)} & \multirow{2}{*}{Mode II(b)} \\
 \cline{2-3}
 ~ &Mode I(a) &Mode II(a) &~ &~ \\
 % Source &Mode I(a) &Mode II(a) &Mode I(b) &Mode II(b) \\
 \hline\hline
 $N_{J/\psi}$ & $0.4^*$ & $0.4^*$ & 0.4 & 0.4 \\
 \hline
 Photon detection & $1.0^*$ & $2.0^*$ & 1.0 & 2.0 \\
 \hline
 $\Lambda$ reconstruction  & $0.7^*$ & $0.7^*$ & 0.7 & 0.7 \\
 \hline
 Kinematic fit & $3.4^*$ & $3.4^*$ & 3.4 & 3.4 \\
 \hline
 Signal model &1.3 &1.6 &2.3 &2.2 \\
 \hline
 Fit range & 2.5 & 2.0 &\dots & 4.4 \\
 \hline
%  Signal Shape &xxx & xxx & xxx \\
%  \hline
 Background shape & 4.1 & 4.6 &\dots & 4.1 \\
 \hline
% $(M^{rec}_{\Lambda \Sigma})^{2}<0.01$ $(\text{GeV}/c^2)^2$ & --- & 1.1 & --- & 2.4 \\
$(M^{\rm rec}_{\Lambda \Sigma})^{2}$  & --- & 1.1 & --- & 2.4 \\
 \hline
 $\chi^{2}_{\rm 2C}$ & --- & 0.5 & --- & 1.4 \\
 \hline
 $\mathcal{B}(\Lambda\to p \pi)$ & $1.6^*$ & $1.6^*$ & 1.6 & 1.6 \\
 \hline\hline
 Total  & \multicolumn{2}{c|}{7.4} & 4.6 & 8.2 \\
 \hline
 \hline 
\end{tabular}
\end{table}

All the above contributions and the uncertainty from
the branching fraction of  $\Lambda\to p\pi^-$ are summarized in Table~\ref{totSys_combine},
where the total systematic uncertainty is given by the
quadratic sum of the individual errors, assuming all sources
are independent. In the calculation of  the total systematic uncertainty of $\mathcal{R}$,  the common systematic uncertainties cancel, $e.g.$, the uncertainties associated
with the total number of $J/\psi$ events, the $\Lambda$ reconstruction, the kinematic fit and the branching fraction of $\Lambda\to p\pi^-$. 
For the decay $\Lambda(1670)\to\gamma\Lambda$, the additive systematic uncertainties related to the fit procedures are not present in Table~\ref{totSys_combine}, as they have been considered in the determination of the upper limit described above.

In summary, with a sample of 10 billion $J/\psi$ events taken with the BESIII detector, we observe the radiative transition $\Lambda(1520)\to\gamma\Sigma^0$ with a statistical significance of $16.6\sigma$ for the first time. The ratio of the branching fractions of $\Lambda(1520)\to\gamma\Lambda$ to $\Lambda(1520)\to\gamma\Sigma^0$ is determined to be $2.88 \pm 0.27 (\text{stat.}) \pm 0.21 (\text{syst.})$, which is consistent with the prediction of flavor SU(3)~\cite{ Landsberg}. Consequently, based on the $\mathcal{B}(\Lambda(1520)\to\gamma\Lambda)$ given by Particle Data (PDG)~\cite{PDG}, the branching fraction of $\Lambda(1520)\to\gamma\Sigma^0$ is measured to be $\mathcal{B}(\Lambda(1520)\to\gamma\Sigma^0)=(2.95\pm0.28 (\text{stat.}) \pm 0.56 (\text{syst.}))\times10^{-3}$. Based on the width of $\Lambda(1520)$ given by PDG~\cite{PDG}, the radiative width of $\Lambda(1520)\to \gamma\Sigma^0$ is measured to be $(47.2\pm4.5 (\text{stat.}) \pm 9.0 (\text{syst.}))$ keV. The result is approximately below one sixth of the prediction of RCQM~\cite{RCQM}, which also shows a significant discrepancy for the $\Sigma^0 \to \Lambda$ transition moment, suggesting that this model may not adequately describe the electromagnetic decay of $\Lambda^*$. Similarly problematic, the partial width of $\Lambda(1520)\to\gamma\Sigma^0$ is 3 to 4 times smaller than the prediction from Algebraic model~\cite{AM}. This model also fails to quantitatively describe the transition rates, which are defined only qualitatively. As a result, the significant discrepancies observed with the predictions from both the RCQM and the Algebraic model strongly indicate that these two models are unlikely to serve as suitable representations. However, our measurements of $\Lambda(1520)\to\gamma\Sigma^0$ partial widths, combined with PDG values~\cite{PDG} for $\Lambda(1520)\to\gamma\Lambda$, demonstrate that only NRQM calculations~\cite{QM3} are currently compatible with experimental data.
Although there are clear discrepancies between our results and other theoretical models, 
the limited statistical precision of our data and incomplete uncertainty quantification in the theoretical models currently preclude definitive assessment of their validity.

Moreover, the $\Lambda(1670) \to \gamma \Sigma^0$ decay is observed with a statistical significance of $23.5\sigma$ for the first time, and the product branching fraction is determined to be $\mathcal{B}(J/\psi \to \bar{\Lambda}\Lambda(1670) \times \mathcal{B}(\Lambda(1670) \to \gamma \Sigma^0) = (5.39 \pm 0.29 (\text{stat.}) \pm 0.44 (\text{syst.})) \times 10^{-6}$. Of interest is the absence of this structure in the  $\gamma\Lambda$ mass spectrum, and the upper limit on the product branching fraction is $\mathcal{B}(J/\psi\to\bar\Lambda\Lambda(1670) \times \mathcal{B}(\Lambda(1670)\to\gamma\Lambda)<5.97 \times {10^{ - 7}}$ at the 90\% confidence level and the corresponding ratio of these two decays is determined to be less than 0.11.  Currently, no well-established theory explains this phenomenon. One possible explanation is that this resonant structure may actually originate from $\Sigma(1670)$ rather than $\Lambda(1670)$. In this scenario, it is reasonable to expect that the decay of $\Sigma(1670)\to\gamma \Lambda$ is suppressed. However, the decay $J/\psi\to \bar\Lambda \Sigma(1670)$ would also be suppressed due to the isospin violation. Given the existing limited statistics, it is challenging to distinguish between the resonant structures of $\Lambda(1670)$ and $\Sigma(1670)$. Since the dominant decay of $\Sigma(1670)$ is $\Sigma(1670)\to\Sigma^0 \pi^0$, future investigations utilizing partial wave analysis of $J/\psi \to \bar{\Lambda} \Sigma^0 \pi^0$ may provide clearer insights into this issue.

The BESIII Collaboration thanks the staff of BEPCII and the IHEP computing center for their strong support. This work is supported in part by National Key R\&D Program of China under Contracts Nos. 2023YFA1606000, 2023YFA1606704; National Natural Science Foundation of China (NSFC) under Contracts Nos. 11635010, 11935015, 11935016, 11935018, 12025502, 12035009, 12035013, 12061131003, 12192260, 12192261, 12192262, 12192263, 12192264, 12192265, 12221005, 12225509, 12235017, 12361141819, 12475089; the Chinese Academy of Sciences (CAS) Large-Scale Scientific Facility Program; CAS under Contract No. YSBR-101; 100 Talents Program of CAS; The Institute of Nuclear and Particle Physics (INPAC) and Shanghai Key Laboratory for Particle Physics and Cosmology; German Research Foundation DFG under Contract No. FOR5327; Istituto Nazionale di Fisica Nucleare, Italy; Knut and Alice Wallenberg Foundation under Contracts Nos. 2021.0174, 2021.0299; Ministry of Development of Turkey under Contract No. DPT2006K-120470; National Research Foundation of Korea under Contract No. NRF-2022R1A2C1092335; National Science and Technology fund of Mongolia; Polish National Science Centre under Contract No. 2024/53/B/ST2/00975; Swedish Research Council under Contract No. 2019.04595; U. S. Department of Energy under Contract No. DE-FG02-05ER41374.

\nocite{*}

\bibliography{paper}% Produces the bibliography via BibTeX.

\end{document}